\documentclass[%
 reprint,
 amsmath,amssymb,
 aps,
]{revtex4-1}

\usepackage{graphicx}
\usepackage{dcolumn}
\usepackage{bm}
\usepackage{textcomp}

\begin{document}


\title{The effect on Kerr comb generation in a clockwise and counter-clockwise mode coupled microcavity}

\author{Shun Fujii}
\affiliation{Department of Electronics and Electrical Engineering, Faculty of Science and Technology, Keio University, Yokohama, 223-8522, Japan}

\author{Atsuhiro Hori}
\affiliation{Department of Electronics and Electrical Engineering, Faculty of Science and Technology, Keio University, Yokohama, 223-8522, Japan}%

\author{Takumi Kato}
\affiliation{Department of Electronics and Electrical Engineering, Faculty of Science and Technology, Keio University, Yokohama, 223-8522, Japan}

\author{Ryo Suzuki}
\affiliation{Department of Electronics and Electrical Engineering, Faculty of Science and Technology, Keio University, Yokohama, 223-8522, Japan}

\author{Yusuke Okabe}
\affiliation{Department of Electronics and Electrical Engineering, Faculty of Science and Technology, Keio University, Yokohama, 223-8522, Japan}%

\author{Wataru Yoshiki}
\affiliation{Department of Electronics and Electrical Engineering, Faculty of Science and Technology, Keio University, Yokohama, 223-8522, Japan}

\author{Akitoshi Chen-Jinnai}
\affiliation{Department of Electronics and Electrical Engineering, Faculty of Science and Technology, Keio University, Yokohama, 223-8522, Japan}

\author{Takasumi Tanabe}
\email{takasumi@elec.keio.ac.jp}
\affiliation{Department of Electronics and Electrical Engineering, Faculty of Science and Technology, Keio University, Yokohama, 223-8522, Japan}


%

\date{\today}

\begin{abstract}
We study the impact of inherent mode coupling between clockwise (CW) and counter-clockwise (CCW) modes on Kerr comb generation in a small whispering-gallery mode microcavity. Our numerical analysis using a coupled Lugiato-Lefever equation reveals the range of the coupling strength in which a soliton pulse can be obtained in the CW direction. It also showed that CCW comb power depends on the coupling strength between the CW and CCW modes. In addition to the simulation, we conducted an experiment to confirm that the power ratio between the CW and CCW comb modes depends on the coupling strength, and the experimental results agree well with the simulation results. This study helps us to understand the relationship between CW and CCW mode coupling and Kerr comb generation, and the effect on soliton formation.
\end{abstract}

\maketitle


\section{Introduction}

Over the past few years, intense research has been undertaken on Kerr combs using whispering gallery mode (WGM) microcavities~\cite{del2007optical,del2011octave,grudinin2009generation,agha2009theoretical,kippenberg2011microresonator,ferdous2011spectral,liang2011generation,wang2012observation,moss2013new,papp2013mechanical,wang2013mid,Coen:13,1347-4065-55-7-072201,Kato:17}. An optical frequency comb in a microcavity, which we call a Kerr comb, has great potential for many applications including precise spectroscopy~\cite{Suh600}, optical clocks~\cite{papp2014microresonator}, microwaves~\cite{Liang2015:high}, and high-speed optical communications~\cite{Pfeifle2014}. The generation of temporal solitons in a microcavity was achieved some time ago, and this enabled the stable and efficient generation of a Kerr comb~\cite{herr2014temporal,yi2015soliton,brasch2016photonic}. In addition to the generation of a bright soliton, a recent study revealed that it is even possible to generate a dark soliton in a normal dispersion regime~\cite{xue2015mode}. This is achieved as a result of local dispersion changes realized using mode coupling~\cite{savchenkov2012kerr,xue2015mode,liu2014investigation,huang2015mode,xue2015normal,Jang:16}.

Mode coupling occurs between different order modes (i.e. $\mathrm{TE_{1}}$-$\mathrm{TE_{2}})$, between different polarization modes (i.e. TE-TM), between two coupled cavities, and between clockwise (CW) and counter-clockwise (CCW) modes, which is known as CW-CCW mode coupling. Such CW-CCW mode coupling is the result of surface scattering by nanoparticles or defects in a WGM microcavity, and it has been observed in microspheres~\cite{weiss1995splitting,kippenberg2002modal,gorodetsky2000rayleigh,PhysRevLett.99.173603}, microrods~\cite{del2013laser} and microtoroids~\cite{PhysRevLett.103.027406,Zhu:10,yoshiki2015observation}. When different modes couple with each other, the resonant spectrum exhibits anti-crossing behavior, and this spectrum split is used to induce a local dispersion modulation, which assists the generation of a comb in a normal dispersion regime~~\cite{savchenkov2012kerr,xue2015mode,liu2014investigation,huang2015mode,xue2015normal,Jang:16} or distorts the comb spectrum~\cite{ramelow2014strong}. Although attention has recently been paid to the influence of mode coupling with other mode families on Kerr comb generation, the impact of CW-CCW mode coupling has received attention only very recently\cite{Yang2017}, despite it being inherent in a small WGM microcavity~\cite{yoshiki2015observation}. 

\begin{figure}[h]
	\centering
	\includegraphics{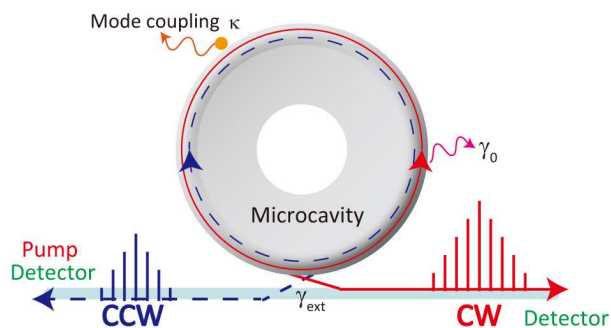}
	\caption{\label{fig:1} Schematic illustration of a CW and CCW coupled cavity. The cavity is pumped in the CW direction, and the comb outputs in both the CW and CCW directions are recorded.}
\end{figure}

In this paper, we study the influence of CW-CCW mode coupling and discuss the formation of a Kerr comb and a soliton in a mode coupled WGM microcavity. For the numerical simulation, we modeled a CW-CCW mode coupled system and investigated Kerr comb generation using a constant or random mode coupling rate with different input powers and coupling strengths. We demonstrated CW and CCW comb formation using a silica toroid microcavity experimentally, and investigated the way in which the CW-CCW coupling affects the power ratio of the generated CW-CCW combs. Figure~\ref{fig:1} shows a schematic illustration of Kerr comb generation in a CW-CCW coupled cavity. 

\section{Modeling of coupled cavity system}
We employed Lugiato-Lefever equations (LLE)~\cite{lugiato1987spatial,Matsko:11,Coen:13,Leo2010,PhysRevA.87.053852} to simulate comb generation. When we consider a single cavity system, the equation is expressed as follows:

\begin{equation}
\begin{split}
\frac{\partial A(\phi,t)}{\partial t} = -\left( \frac{\gamma_{\mathrm{tot}}}{2} +i \delta_0 \right)A +i \frac{D_2}{2} \frac{\partial^2 A}{\partial \phi^2} + i g |A|^2 A  \\+ \sqrt{\gamma_{\mathrm{ext}}}A_{\mathrm{in}}, \label{Eq.1}
\end{split}
\end{equation}
where $\phi$ is the azimuthal angle along the circumference of the cavity, $t$ is the time that describes the evolution of the field envelope, $\gamma_{\mathrm{tot}}$ is the cavity decay rate, $\delta_0=\omega_0-\omega_p$ is the detuning of the light frequency from the resonance, $D_2$ is the second-order dispersion, $g$ is a nonlinear coefficient, and $A_{\mathrm{in}}=\sqrt{P_{\mathrm{in}}/\hbar \omega_p}$ is the input field. Note that the formula for the cavity decay rate, quality factor, and resonance frequency $\omega_0$ is given as $\gamma_{\mathrm{tot}}=\omega_0/Q=\gamma_0+\gamma_{\mathrm{ext}}$, where $Q_0=\omega_0/\gamma_0$ is the intrinsic $Q$ factor and $Q_{\mathrm{ext}}=\omega_0/\gamma_{\mathrm{ext}}$ is the external $Q$ factor. The nonlinear coefficient $g$ is given as $\hbar \omega_0^2 c n_2 /(n_0^2 V_\mathrm{eff})$, where  $n_2$ is the nonlinear refractive index, $c$ is light speed in a vacuum and $V_{\mathrm{eff}}=A_{\mathrm{eff}}L$ is the effective mode volume of the cavity (where $A_{\mathrm{eff}}$ is the effective mode area and $L$ is the cavity length). In this work, we model CW-CCW mode coupling and so we added a mode coupling term to the Lugiato-Lefever equations~\cite{Yi2017}:

\begin{equation}
\begin{split}
\frac{\partial A(\phi,t)}{\partial t} = -\left( \frac{\gamma_{\mathrm{tot}}}{2} +i \delta_0 \right)A +i \frac{D_2}{2} \frac{\partial^2 A}{\partial \phi^2} + i g |A|^2 A  \\+ i \frac{\kappa_\mu}{2}B + \sqrt{\gamma_{\mathrm{ext}}}A_{\mathrm{in}}, \label{Eq.2}
\end{split}
\end{equation}
\begin{equation}
\begin{split}
\frac{\partial B(\phi,t)}{\partial t} = -\left( \frac{\gamma_{\mathrm{tot}}}{2} +i \delta_0 \right)B +i \frac{D_2}{2} \frac{\partial^2 B}{\partial \phi^2} + i g |B|^2 B \\+ i \frac{\kappa_\mu}{2}A, \label{Eq.3}
\end{split}
\end{equation}
where $\kappa_\mu$ is the linear coupling coefficient between modes $A$ and $B$, which correspond to the CW and CCW modes, respectively. The difference between this coupling coefficient and that in Refs.~\cite{Yi2017,liu2014investigation} results from different definitions of coupling rate. In this work, $\kappa/2\pi$ is the width of resonance splitting~\cite{yoshiki2015observation}. By setting the $\kappa_\mu$ value for each mode, we can calculate the linear and nonlinear effects in a CW-CCW coupled cavity. The simulations are implemented using the split-step Fourier method with 256 frequency modes. In addition to frequency and time domain analysis, we monitored the average intracavity power. This analysis helps us to understand the dynamics of comb generation in a mode coupled cavity system. The third terms in Eqs.~(\ref{Eq.2}) and (\ref{Eq.3})  on the right-hand side describe the Kerr effects (self-phase modulation). In this model, we neglected the cross-phase modulation (XPM) between counter-propagating modes because the XPM interaction is very weak and can be neglected for short pulses~\cite{agrawal2007nonlinear}. However, even when we considered the nonlinear coupling between counter-propagation waves, we confirmed that the effect is small (see Appendix).

\section{\label{sec:num} Numerical simulation of Kerr comb formation}
\subsection{Without mode coupling}

Before calculating the comb generation in the CW-CCW coupled model, we simulated a Kerr comb without mode coupling to confirm whether or not the selected parameters were appropriate and the results comparable to those obtained with the CW-CCW coupled model that we describe later. In this calculation, we consider the following parameters for a silica toroid microcavity whose major and minor diameters are 50 and 7~\textmu m, respectively: input wavelength $\lambda_p=1542$~nm, refractive index $n=1.44$, nonlinear refractive index $n_2=2.2\times10^{-20}$~$\mathrm{m^2/W}$, $Q_0=2\times10^7$, $Q_{\mathrm{ext}}=2\times10^7$ and  $A_{\mathrm{eff}}=5$~\textmu $\mathrm{m^2}$. The cavity free-spectral range (FSR) $D_1/2\pi$ is 1350~GHz, and the second-order dispersion $D_2/2\pi=23.8$~MHz (anomalous group-velocity dispersion). These values are calculated with the finite element method (FEM). Note that these parameters are for the second-order TE mode in a silica toroid.

\begin{figure}[h]
	\centering
	\includegraphics{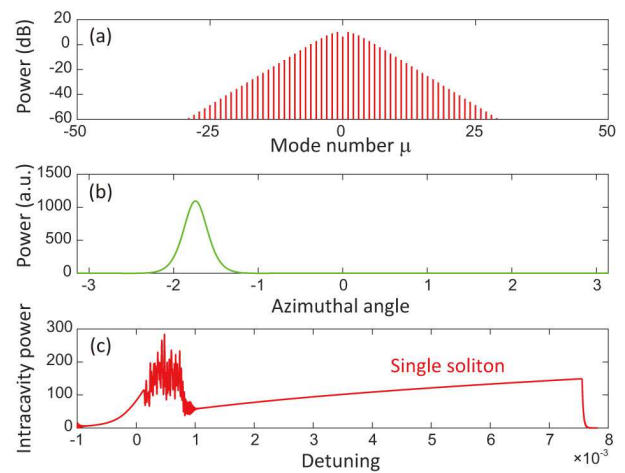}
	\caption{\label{fig:2} (a,b) Simulated optical spectrum and temporal waveform, respectively. (c) Average intracavity power versus detuning during laser scanning from blue to red.}
\end{figure}

Figure~\ref{fig:2}(a) and (b) are the calculated comb spectrum and the time domain waveform, respectively, with 100~mW pumping. As shown in Fig.~\ref{fig:2}(c), the intracavity intensity exhibited chaotic behavior and then became stable as we swept the input laser. As already reported in many previous studies \cite{herr2014temporal,Guo2017}, this is typical soliton generation behavior (e.g. soliton step) in an anomalous dispersion regime. The appearance of such behavior in a coupled cavity system is an indication as to whether or not modal coupling affects soliton generation. 

\subsection{Constant coupling with all modes}

\begin{figure}[!b]
	\centering
	\includegraphics{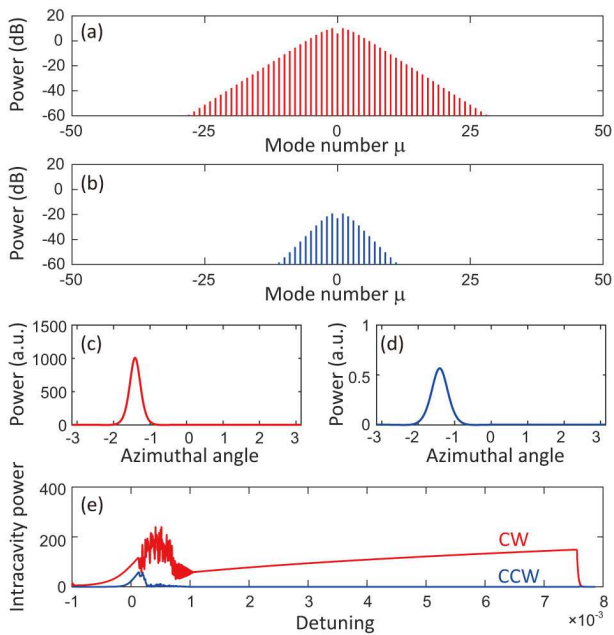}
	\caption{\label{fig:3} Simulated optical spectrum (a,b) and temporal waveform (c,d) of CW (red) and CCW (blue) directions when the CW-CCW mode coupling is the same for all longitudinal modes with a constant $\Gamma=1.0$, $P_{\mathrm{in}}=100$~mW. (e) Average intracavity power and normalized detuning. Red and blue lines represent the CW and CCW directions, respectively.}
\end{figure}

Our main interests are as follows; whether or not soliton formation with the CW mode is affected by CCW mode coupling, the mechanism of CCW comb generation and whether or not CW and CCW soliton pulses are obtained simultaneously in spite of unidirectional pumping. To answer these questions, first, we assume a constant coupling parameter for all modes, which corresponds to CW-CCW mode coupling with the same coupling strength. The mode coupling parameter $\Gamma=\kappa_\mu/\gamma$ represents the coupling strength of each mode, and the other parameters are the same as in Fig.~\ref{fig:2}. The threshold power of parametric four-wave mixing (FWM) is described in Ref.~\cite{herr2012universal}; therefore the calculated threshold is about 1~mW. The CCW mode experiences parametric oscillations when the FWM gain exceeds the cavity loss.

Figure~\ref{fig:3} shows the calculated results with constant coupling $\Gamma=1.0$ and input power $P_{\mathrm{in}}=100$~mW. In this condition, the obtained CW comb spectrum and intracavity intensity are almost the same as those in Fig.~\ref{fig:2}, which indicates that the CW comb is not disturbed by mode coupling with $\Gamma=1.0$. However, the intracavity power of the CCW mode did not exhibit typical soliton formation behavior. Although the intracavity power of both the CW and CCW modes increased until the detuning reached $0.2\times 10^{-3}$, the power of the CCW mode decreased significantly when the CW mode generated a pulse ($> 1\times 10^{-3}$). We confirmed that the CCW mode receives the power of the CW comb by fixed coupling in the spectrum domain. As a result, the spectrum and the time domain waveform of the CCW mode are similar to those of the CW mode; however the peak power is orders of magnitude smaller than that of the CW mode. 

\begin{figure}[!b]
	\centering
	\includegraphics{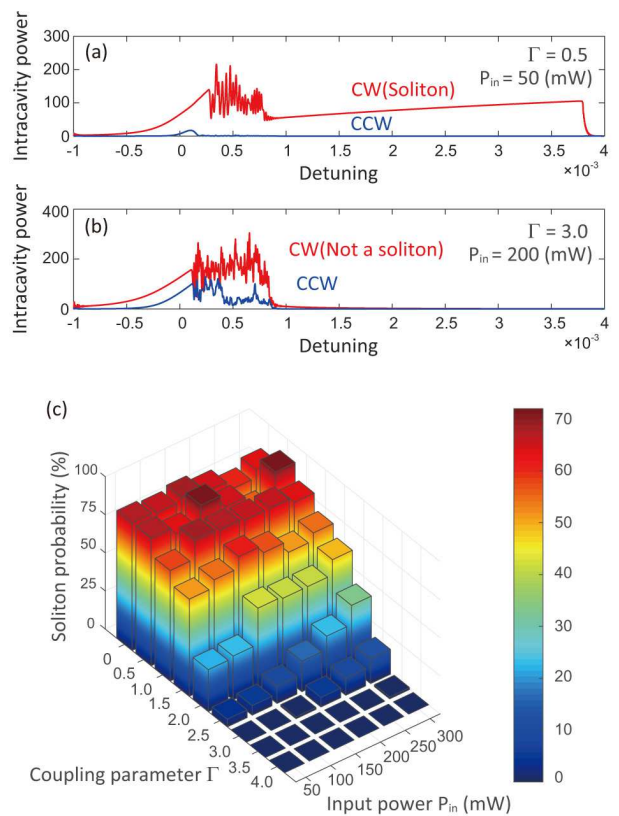}
	\caption{\label{fig:4} (a) Average intracavity power with $\Gamma=0.5$, $P_{\mathrm{in}}=50$~mW. CW direction forms a soliton. (b) Simulated results with $\Gamma=3.0$, $P_{\mathrm{in}}=200$~mW. In this case, neither the CW nor CCW mode can achieve a soliton state. (c) Numerical investigation of probability of soliton formation in the CW direction. In each case, the calculations were performed 100 times.}
\end{figure}

To undertake further analysis, we changed the coupling strength and input power parameters. Figure~\ref{fig:4}(a) and (b) show the intracavity intensity with $\Gamma=0.5$, $P_{\mathrm{in}}=50$~mW and $\Gamma=3.0$, $P_{\mathrm{in}}=200$~mW, respectively.  Figure~\ref{fig:4}(c) shows a numerical investigation of the probability of soliton generation. This investigation allows us to know the range in which soliton formation in the CW mode is unaffected by coupling. We obtained only two types of result; one was a soliton state in the CW direction and the other was a no soliton state in either direction. With a small $\Gamma$ value, the CW mode exhibits a stable soliton state. However, with a $\Gamma$ value of $>2.5$, the probability suddenly decreases.  When $\Gamma>3.0$, the CW mode could not be transformed into a soliton state. From this result, we found that the mode coupling strength affects the probability of soliton formation in the CW mode. Only very strong coupling induces competition between the intracavity powers of the CW and CCW modes and disturbs Kerr comb formation in the CW mode. In other words, the coupling between the CW-CCW modes will not affect the generation of a Kerr comb in the CW direction as long as the coupling is within the practical strength range.

We also considered a case where scattering occurs at a different position at each roundtrip by adding a phase term $\kappa_\mu=\kappa'_\mu e^{i \psi_\mu}$ to the coupling term, where $\psi_\mu$ is randomly changed at each time step. We found that the CW mode is not affected by the random phase, but the spectrum and the waveform at CCW are affected. The spectrum and the temporal waveform vary at each roundtrip, and the CCW output is unstable. This indicates that the output in the CCW direction is unstable when the scattering occurs at a random position in the cavity. However, as we will show later, the spectrum in the CCW direction is stable in the experiment, which suggests that the CW-CCW mode coupling is the result of scattering at nanoparticles and defects in a WGM cavity and does not changethe position in time.

\subsection{Random coupling with all modes}
\begin{figure}[h]
	\centering
	\includegraphics{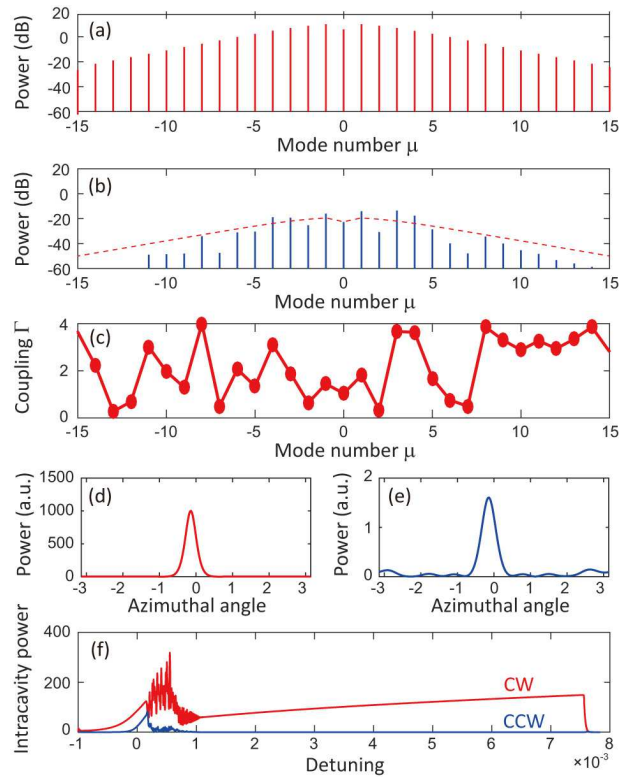}
	\caption{\label{fig:5} (a,b) Simulated optical spectrum of the CW and CCW modes when the coupling strength is randomly set with $\Gamma=0$ to 4.0 as shown in (c). The red dashed line in (b) shows the envelope of the CW comb spectrum in (a). (d,e) Simulated temporal waveforms for CW (red) and CCW (blue) directions. (f) Average intracavity power and normalized detuning of CW and CCW modes.}
\end{figure}

Thus far we have assumed constant coupling, however in fact the coupling strength is not always the same for all modes. This has been observed experimentally and can be explained by the presence of multiple scattering points~\cite{zhu2010chip}. We also confirm the way in which the difference between the coupling parameters of each longitudinal mode affects Kerr comb generation. Figure~\ref{fig:5} shows the results with $\Gamma=0\sim4.0$ and $P_{\mathrm{in}}=100$~mW. The $\Gamma$ value of each mode is randomly selected from 0 to 4.0 by a pseudorandom number algorithm whose average value is about 2.0. Calculated spectra of the CW and CCW modes are shown in Fig.~\ref{fig:5}(a) and (b), respectively. We show only 30 modes to make it easy to compare the power difference. The CCW mode experienced random coupling $\Gamma$ from 0 to 4.0 as shown in Fig.~\ref{fig:5}(c). In particular, the CCW comb intensities of modes $\mu=-7$, 2 and 7 are weaker than those of the other modes depending on the coupling strength $\Gamma$. On the other hand, the intensity of some combs (i.e.~$\mu=3, 4$) is correspondingly strong. Even when a CW mode is in a soliton state (Fig.~\ref{fig:5}(a) and (f)) or in a Turing pattern state (not shown), the intensity of the CCW mode is weak and the waveform is not smooth (Fig.~\ref{fig:5}(e)). These results show that the CCW comb teeth are directly affected by the coupling strength and could not be compensated with the nonlinear effect. On the other hand, the CW mode generates a soliton pulse in spite of random coupling as shown in Fig.~\ref{fig:5}(d). 

\section{Experimental measurement}

In this section, we describe an experimental demonstration of Kerr comb generation with a CW-CCW mode coupled silica toroid microcavity. The numerical results in Fig.~\ref{fig:5}(b) and (c) show that a CCW comb clearly has a relationship with the coupling strength of each mode. From this point of view, when we observe the comb spectra for both directions, the difference in coupling strength should appear in the frequency spectrum. 

We fabricated a silica toroid microcavity using photolithography, $\mathrm{SiO_2}$ etching, $\mathrm{XeF_2}$ gas etching and a $\mathrm{CO_2}$ laser reflow process \cite{armani2003ultra}. The fabricated microcavity had a diameter of 45~\textmu m. The diameter was carefully selected, because the dispersion of a microcavity with too small a diameter is normal, whereas an anomalous dispersion is needed to generate a Kerr comb \cite{hansson2013dynamics,matsko2012hard}. We used the second-order mode because it exhibits an anomalous dispersion. The mode is selectively excited by adjusting the position of a tapered fiber~\cite{Lin:10,Fujii:17}. It should be noted that a WGM microcavity with a small diameter exhibits strong CW-CCW mode coupling.

\begin{figure}[b]
	\centering
	\includegraphics{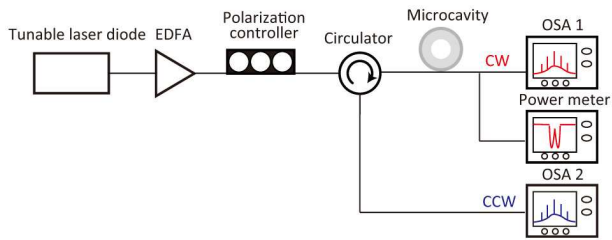}
	\caption{\label{fig:6} Experimental setup used for measuring combs and transmission spectra.}
\end{figure}

The setup we used for our experimental measurement is shown in Fig.~\ref{fig:6}. The light from a tunable CW laser operating at 1548~nm was input into the cavity through an optical circulator. The amplified input power of $\sim$300~mW and the generated comb were observed with an optical spectrum analyzer (OSA). A CCW comb could be also analyzed through a circulator with another OSA. We employed a tapered fiber with a diameter of 1~\textmu m to couple the light into the microcavity. 

\begin{figure}[b]
	\centering
	\includegraphics{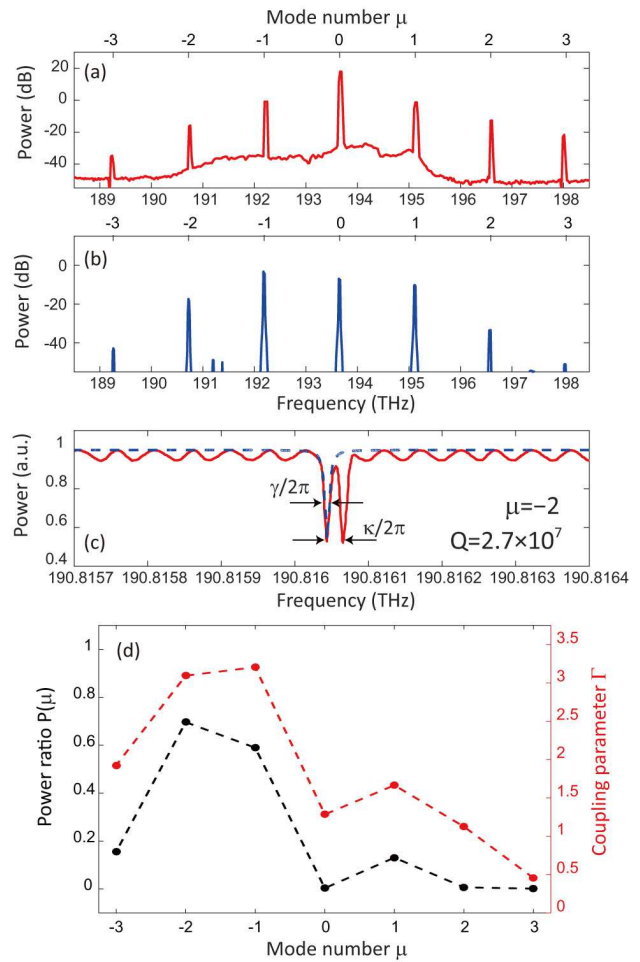}
	\caption{\label{fig:7} (a) Measured CW comb spectrum. (b) Measured CCW comb spectrum. (c) Transmittance spectrum of a mode ($\mu=-2$). $Q$ is $2.7\times10^7$ and the splitting width $\kappa/2\pi$ is 22~MHz. (d) Relationship between the normalized power ratio of CW to CCW and the coupling parameter $\Gamma$.}
\end{figure}

First, we simultaneously measured the combs generated in the CW and CCW directions by using a circulator at the input port. The results are shown in Figs.~\ref{fig:7}(a) and (b), where we observe clear comb spectra in both directions. It should be noted that the envelope of the CW comb is triangular, which suggests that the comb in the CW direction is in a low-noise initial state, which is known as a Turing pattern. On the other hand, the CCW direction is stable with respect to time but produces an irregular spectrum shape. We measured the transmittance spectrum of the longitudinal lines precisely and characterized the mode splitting. The result for one of the lines is shown in Fig.~\ref{fig:7}(c), where we observe a clear mode splitting spectrum with a splitting width of $\kappa/2\pi=22$~MHz. The coupling strength differs for different modes, but falls within the 10 to 50~MHz range. It should be noted that the typical measured linewidth implies $Q\sim1\times10^7$. Although the output from a charged cavity generally exhibits Rabi-like oscillation~\cite{yoshiki2015observation}, the output is a stable continuous wave when the mode is continuously sustained by FWM gain. Thus we confirmed that the generated comb was stable even for a system with a coupled CW-CCW mode. 

Next, we compared the power ratio of the CW and CCW modes ($P_{\mu}$) with the strength of the CW-CCW coupling ($\Gamma=\kappa/\gamma$). We calculated $P_{\mu}$ from the CW and CCW powers of the modes by using Figs.~\ref{fig:7}(a) and (b). Moreover, we measured the transmission spectrum of each longitudinal line and carefully obtained the splitting width $\kappa/2\pi$, as shown in Fig.~\ref{fig:7}(c). Here, we calculated the power ratio $P_{\mu}$ by dividing the CCW peak power by the CW peak power. The $P_{\mu}$ and $\Gamma$ results for each mode are summarized in Fig.~\ref{fig:7}(d), where the two lines show a strong correlation, which indicates that the CCW comb is generated by the scattering of each CW comb component. This in turn indicates that the FWM process is dominant in the CW direction, and nonlinear mixing rarely occurs in the CCW direction. We confirmed that the coupling strength of each mode is a dominant factor in the CCW direction, and the trend of these experimental results agrees with the numerical simulation shown in Fig.~\ref{fig:5}.

As already reported in previous work \cite{ramelow2014strong,grudinin2013impact}, the strong crossing of modes influences the comb spectrum since the resonance positions of longitudinal modes are especially important for the phase-matching condition. However, in our experiment, CW-CCW modal coupling is limited, therefore the comb spectrum does not seem to suffer any serious effects.

These results indicate that it will be possible to determine the coupling strength of each longitudinal mode by employing a power ratio measurement with FWM, which is a much easier and more efficient way to measure the coupling strength than to measure the spectrum splitting of each longitudinal line individually.

When we measured the transmission spectrum, we observed only that in the CW direction because it is sufficient for determining the split width. To measure the coupling, we need to precisely measure the transmittance spectrum in the CW direction and resolve the spectrum splitting, which is often not an easy task, especially when we are interested in more than one longitudinal line. On the other hand, if we observe the generated comb in both the CW and CCW directions simultaneously, we might be able to obtain information about the CW-CCW mode coupling in the broad wavelength regime immediately, without employing an ultraprecise transmittance spectrum measurement. Therefore, an understanding of the relationship between comb generation and CW-CCW mode coupling may also be advantageous as regards the CW-CCW coupling measurement.

\section{Conclusion}
In conclusion, we demonstrated Kerr comb generation in a CW and CCW mode coupled microcavity both experimentally and theoretically. In the numerical simulation, we developed a model with which to study a CW-CCW mode coupled system based on a Lugiato-Lefever equation. The calculation result suggested that CW-CCW coupling will not degrade the CW comb and soliton formation as long as the system has a sufficiently anomalous dispersion and weak coupling. In the experiment, we observed the contribution of linear scattering to the generation of a comb-like spectrum in the CCW direction. The splitting width and the ratio of the generated CW-CCW comb components showed good agreement thus suggesting that this method can even be used to determine the coupling strength of each longitudinal mode simultaneously. We hope that this work will contribute to a better understanding of the recent study of Kerr combs using mode coupling.

\setcounter{section}{0}
\renewcommand{\thesection}{\Alph{section}}
\setcounter{equation}{0}
\renewcommand{\theequation}{\Alph{section}.\arabic{equation}}
\setcounter{figure}{0}
\renewcommand{\thefigure}{\Alph{section}.\arabic{figure}}
\setcounter{table}{0}

\section*{Appendix}
\appendix
\section{Effect of nonlinear couplings between counter-propagating modes}
The coupling between the CW and CCW modes is small, because the symmetry breaking of resonance is induced by cross phase modulation (XPM) between two counter-propagating lights~\cite{DelBino2017}.  Since the nonlinear induced symmetry breaking, which induces the frequency offset between CW and CCW resonant modes, occurs before the generation of FWM, it suppresses the coupling between them. However, there is a possibility that XPM and FWM between counter-propagating modes affects the waveforms as observed in previous studies~\cite{Firth:88, Yang2017,Yariv:77}.  Hence, here we consider the effect of these effects on Kerr comb generation in a counter-propagating model. We followed the model described in Refs.~\cite{Yang2017, Lin:16, dong2016unified}, and added a nonlinear coupling term to the coupled Lugiato-Lefever equations.

\begin{equation}
\begin{split}
\frac{\partial A(\phi,t)}{\partial t} = -\left( \frac{\gamma_{\mathrm{tot}}}{2} +i \delta_0 \right)A +i \frac{D_2}{2} \frac{\partial^2 A}{\partial \phi^2} \\+ i g (|A|^2 +2|B|^2)A  + i \frac{\kappa_\mu}{2}B + \sqrt{\gamma_{\mathrm{ext}}}A_{\mathrm{in}}, \label{Eq.A1}
\end{split}
\end{equation}

\begin{equation}
\begin{split}
\frac{\partial B(\phi,t)}{\partial t} = -\left( \frac{\gamma_{\mathrm{tot}}}{2} +i \delta_0 \right)B +i \frac{D_2}{2} \frac{\partial^2 B}{\partial \phi^2} \\+ i g (|B|^2 +2|A|^2)B  + i \frac{\kappa_\mu}{2}A, \label{Eq.A2}
\end{split}
\end{equation}

\begin{figure}[!h]
	\centering
	\includegraphics{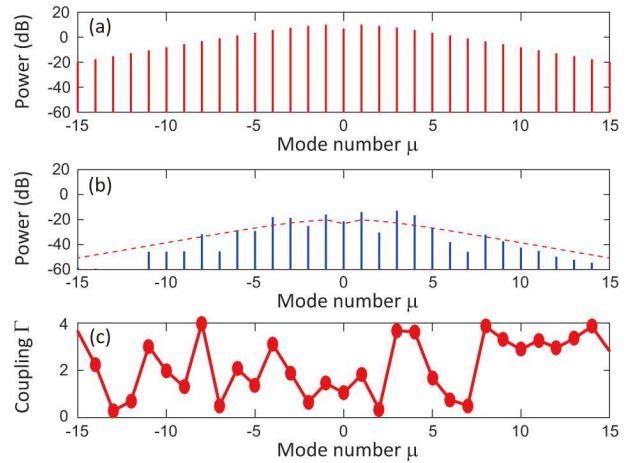}
	\caption{\label{fig:8} (a,b) Simulated optical spectrum of the CW and CCW modes with XPM terms when the coupling strength is randomly set as in Fig.~\ref{fig:5} as shown in (c). The dashed red line in (b) shows the envelope of the CW comb spectrum in (a).}
\end{figure}

We performed calculations with Eqs.~(\ref{Eq.A1}) and (\ref{Eq.A2}), using the parameters described in Section~\ref{sec:num}. Figure~\ref{fig:8} shows the results we obtained with random coupling.  The calculated spectra of the CW and CCW modes are shown in Fig.~\ref{fig:8}(a) and (b), and the coupling rate is shown in Fig.~\ref{fig:8}(c). There is almost no difference between these results and those shown in Fig.~\ref{fig:5}, where no nonlinear coupling is taken into account.  This confirms that the effect of the nonlinear coupling between counter propagating waves is limited in our case.

\section*{Funding}
This work was supported by the Japan Society for the Promotion of Science (JSPS) (KAKEN $\#$15H05429).

\section*{Acknowledgments}
We thank Yanne K. Chembo for helpful discussions and suggestions.

\end{document}